\documentclass[prb,twocolumn,aps,showpacs]{revtex4}
\usepackage{graphicx}
\usepackage{amsmath}
\usepackage{widetext}
\begin{document}

\title{Rabi-vibronic resonance with large number of vibrational quanta}
\author{R. Glenn  and M. E. Raikh }
\affiliation{Department of Physics, University of Utah, Salt Lake City, UT 84112}

\begin{abstract}
We study theoretically the Rabi oscillations of a resonantly driven
two-level system linearly coupled to a harmonic oscillator
(vibrational mode) with frequency, $\omega_0$. We show that for weak
coupling, $\omega_p \ll \omega_0$, where $\omega_p$ is the polaronic
shift, Rabi oscillations are strongly modified in the vicinity of the
Rabi-vibronic resonance  
$\Omega_R = \omega_0$, where $\Omega_R$ is the
Rabi frequency.
The width of the resonance is $(\Omega_R-\omega_0) \sim
\omega_p^{2/3} \omega_0^{1/3} \gg \omega_p$.  Within this domain of
$\Omega_R$ the actual frequency 
of the Rabi oscillations
exhibits a bistable behavior as a function of $\Omega_R$. Importantly,
within the resonant domain, the oscillator is highly excited,
which allows  to treat it classically.
\end{abstract}
\maketitle

\section{Introduction}
\begin{figure}
\begin{center}
\includegraphics[scale=.5]{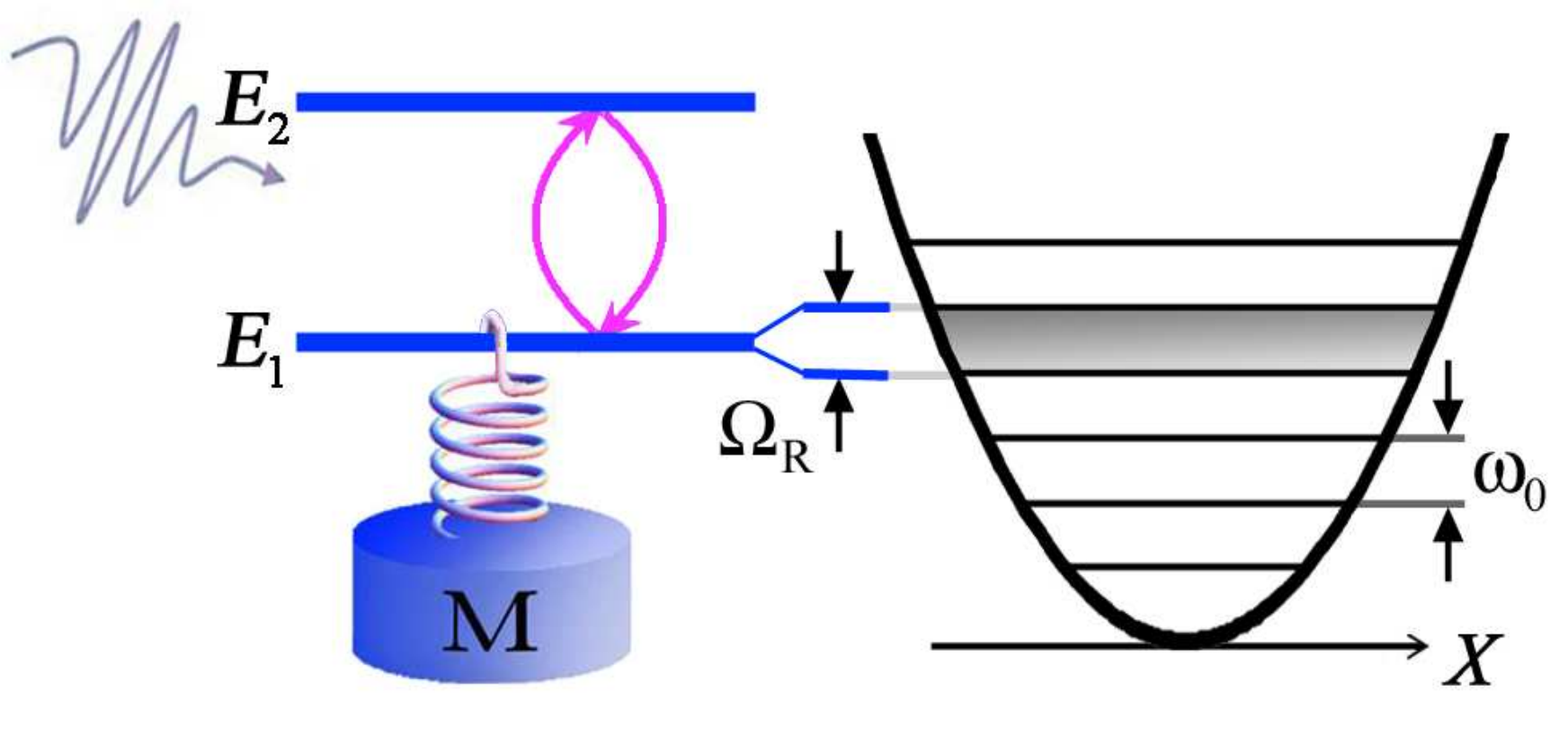}
\end{center}
\caption{A schematic illustration of the system under consideration. Two-level system is driven by near-resonant light, $\omega_{12} \approx (E_2 -E_1)$. The level $E_1$ is linearly coupled to a classical oscillator with frequency $\omega_0$. The Rabi oscillations are strongly modified when $\omega_0$ is close to $\Omega_R$, where $\Omega_R$ is the Rabi frequency.}
\label{twolevelsystem}
\end{figure}

Coupling to environment tends to damp the Rabi oscillations\cite{Rabi} of a
resonantly driven two-level system. Usually, the environment is viewed as
medium with continuous spectrum of modes.
Less common is the
situation when environment possesses a single or several well-defined
frequencies. For concreteness we will consider the situation depicted
in Fig.~\ref{twolevelsystem} when the lower level of the two-level
system is coupled to an oscillator (a mass, $M$, and a spring),
 which represents a single
vibrational mode. Obviously,  coupling to the oscillator
 has a strong effect on  the Rabi oscillations in the regime
of the vacuum Rabi splitting\cite{vacuumRabi} when the oscillator frequency,
$\omega_0$, is close to the transition frequency, $\omega_{12}$.
It is less obvious what effect the coupling to the
oscillator will have on the Rabi oscillations when $\omega_0$ is
much smaller than $\omega_{12}$ and is comparable to the Rabi frequency,
$\Omega_R$.
One can  argue on
physical grounds that the effect of coupling on the Rabi oscillations
will be strong in the vicinity of the condition, $\omega_0 \approx
\Omega_R$, which we  dub {\it Rabi-vibronic resonance}. Indeed,
consider the Hamiltonian
\begin{equation}
\label{Hamiltonian}
H=\mu \hat{X}\hat{n}_{1},
\end{equation}
describing the linear coupling. Here $\hat{X}=\frac{1}{\sqrt 2}(b^{\dagger}+b)$ is the  operator of
the oscillator displacement, $b^{\dagger}$ is a creation operator of the vibrational
quantum, $\hat{n}_1$ is the occupation of the level $E_1$,
and $\mu=\bigl(2M\omega_0^3\bigr)^{1/2}\lambda$,
where $\lambda$ is a
dimensionless coupling constant.
In the course of the  Rabi oscillations  the average $\hat{n}_1$ changes with time as
\begin{equation}
\label{withtime}
n_1(t)=\frac{1}{2}\bigl(1+\cos\Omega_Rt\bigr).
\end{equation}
Then at  $\Omega_R\approx \omega_0$,
the second term in Eq. (\ref{withtime}) 
gives rise to  a resonant driving force acting on the oscillator.
In turn, the strongly driven oscillator
provides a resonant feedback
\cite{shekhterfeedback,feedback}
 on the two-level system. Thus, as
$\Omega_R$, which is  proportional to the ac  field  driving the two-level system,
increases,  we expect the Rabi oscillations to be strongly modified
near the resonant condition.  

Among possible experimental realizations of the situation Fig.~\ref{twolevelsystem} 
 is a suspended carbon nanotube in inhomogeneous electric field, 
which creates a confinement for an exciton\cite{nanotube1,nanotube2,nanotube3,nanotube4}.
 Localized exciton  can be viewed as a two-level system. Bending
modes having discrete frequencies due to finite nanotube length,   
can be viewed as oscillators with very low friction. While a typical transition frequency 
in such a system is\cite{nanotubeAVO,nanotubeCAPAZ} $\omega_{12}\sim 10^{15}$Hz,   
the oscillator frequency\cite{nanotube3} is much smaller $\omega_0 \sim 4.3\cdot 10^{9}$Hz.
Resonant condition can be achieved by adjusting the illumination intensity. 

Another area in which the situation Fig.~\ref{twolevelsystem} is  relevant,  is the  cavity 
QED\cite{QEDGirvin}, where a two-level system is realized in the form of
a  superconducting qubit, while the oscillator is a  LC-circuit. Majority of experimental
and theoretical studies in this field are focused on the strong coupling in the domain 
 $\omega_0\approx\omega_{12}$. However,  in experiments  Refs.~[\onlinecite {QEDIlichev,QEDShnyrkov}]
an ac  driven superconducting qubit was  coupled to a "slow" LC-oscillator tuned 
to $\Omega_R$. It was observed that the noise spectrum
of the oscillator exhibits a Lorentzian peak\cite{QEDSmirnov} as a function of $\Omega_R-\omega_0$.
In theoretical papers\cite{Shnirman,Shnirman1,OnlyOscillatorDriven}
initiated by the experiment Ref.~\onlinecite{QEDIlichev} collective 
motion of the oscillator coupled to a qubit was studied within the density
matrix formalism,  and both subsystems were treated 
quantum-mechanically. In view of complexity of this description, 
final results were obtained numerically for particular values of a 
coupling parameter, $\lambda$. A notable finding of 
Refs.~[\onlinecite{Shnirman,Shnirman1,OnlyOscillatorDriven}] 
is that, in the vicinity of the condition $\omega_0\approx \Omega_R$,
collective Rabi-vibronic motion  becomes {\em bistable}.

There are still several basic questions to be answered, among which: 

{\em (i)} how the
frequency, $s$, of the collective oscillations depends on
 $\lambda$?

{\em (ii)} what is the width of the resonance, i.e., the domain
 $\delta_0=\Omega_R-\omega_0$ of
the Rabi frequencies where Rabi oscillations are modified
due to coupling?

{\em (iii)} how the decay of the Rabi oscillations depends 
on the oscillator friction?

 The above questions are studied in the present paper. Our main
finding is that the width of the Rabi-vibronic resonance is small for
weak coupling, namely, 
\begin{equation}
\label{delta0}
\delta_0 = \lambda^{4/3}\omega_0=\omega_p^{2/3} \omega_0^{1/3}\ll \omega_0,
\end{equation}
where $\omega_p=\lambda^2\omega_0$ is the polaronic shift.
Eq. (\ref{delta0}) suggests that, while $\delta_0$ is much smaller than
$\omega_0$, it is much bigger than $\omega_p$.
Most importantly,  Eq. (\ref{delta0}) guarantees that,  
in the resonant domain $(\Omega_R-\omega_0) \sim \delta_0$,
the oscillator  is highly excited and can be treated as {\em classical}.
This allows the analytical description of the resonance. In this regard,
the situation we consider, two-level system coupled to a classical oscillator,
is  similar to the  {\em Rabi resonance} 
considered in Refs.~[\onlinecite{cubicIlichev,IlichevRabi}],
where two-level system was driven 
by {\em two classical fields}: one with 
frequency close to $\omega_{12}$  and one with frequency close
to $\Omega_R$.

We will see that in the domain $(\Omega_R-\omega_0) \sim \delta_0$, 
the frequency $s$ of  collective oscillations differs from  $\Omega_R$
also by $\sim\delta_0$. Bistable behavior of  the dependence $s(\Omega_R)$
emerges naturally within  our approach; the frequency jumps between two stable
regimes are also $\sim\delta_0$. In addition, in the present paper we  study how 
the Rabi-vibronic resonance depends on detuning, $\Delta$, of the driving 
frequency from $\omega_{12}$, on intrinsic anharmonicity of the oscillator, 
and how the modified Rabi oscillations decay with time due to relaxation
of the  two-level system and due to friction in the oscillator.

\section{Basic equations}

We first assume that the displacement, $X(t)$, is a classical
variable and will later  check this assumption. The equation of motion
for $X(t)$ reads
\begin{equation}
\label{EOMx}
\ddot{X}+ \gamma \dot{X}+\omega_0^2X = \frac{\mu}{2M}\bigl(1-w\bigr),
\end{equation}
where $w=1-2n_1$ is the population inversion, and $\gamma$ is the friction in  the oscillator.
The equation for $w(t)$ and nondiagonal elements of the density matrix have the form of the
optical Bloch equations
\begin{equation}
\label{wdensity}
\dot{w}(t) = -\Omega_R v -\Gamma\bigl(1+w\bigr),
\end{equation}
\begin{equation}
\label{udensity}
\dot{u}(t) =-\Bigl( \Delta -\mu X(t)\Bigr) v -\frac{\Gamma}{2} u,
\end{equation}
\begin{equation}
\label{vdensity}
\dot{v}(t) =\Bigl( \Delta -\mu X(t)\Bigr)u+  \Omega_R w -\frac{\Gamma}{2} v,
\end{equation}
where $\Gamma$ is the relaxation rate of the excited state.
Note that, while the oscillator is driven by $w(t)$, it exercises a feedback on
the two-level system via   $u(t)$ and $v(t)$.

We require that the level $E_1$ at $t=0$ is occupied while the level $E_2$ is empty,
 i.e.,  $w(0)=-1$.  We also assume that the dipole moment and dipole current are initially zero,
 leading to $v(0)=0$ and $u(0)=0$, respectively.
 From Eq. (\ref{wdensity}), we see that the initial conditions
 for $v$ and  $w$ require that $\dot{w}(0)=0$.
\section{Modified Rabi oscillations}
\subsection{Oscillation frequency}

The system  Eqs. (\ref{wdensity})-(\ref{vdensity}) can be reduced to two coupled
equations by excluding $v(t)$ and expressing $u(t)$ in terms of $w(t)$. Then one gets
\begin{equation}
\label{wrabi}
\ddot{w} +   \frac{3}{2}\Gamma \dot{w}+\Bigl(\Omega_R^2+\frac{\Gamma^2}{2}\Bigr)w+\frac{\Gamma^2}{2} =
-\Omega_R\Bigl(\Delta- \mu X(t)\Bigr) u(t),
\end{equation}
\begin{equation}
\label{urabi}
 u(t)=-\!\!\int_0^t\!\frac{ dt^{\prime}}{\Omega_R}e^{\Gamma(t^{\prime}-t)/2}
 \bigl(\mu X(t^{\prime}) - \Delta \bigr) \Bigl [ \dot{w}(t^{\prime}) + \Gamma\bigl (1+ w\bigr) \Bigr]\!.
\end{equation}

 We start from the simplest case, $\Gamma \rightarrow 0$, $\Delta \rightarrow 0$, $\gamma \rightarrow 0$, 
and search for a solution of the system Eqs. (\ref{EOMx}), (\ref{wrabi}), and (\ref{urabi}), in the form 
$w(t) = - \cos s t$.
Substituting this form into  Eq. (\ref{EOMx}) we find the displacement
\begin{equation}
\label{sol rabi x}
X(t) = \frac{\mu \cos s t }{2M\bigl(\omega_0^2 -s^2\bigr)}.
\end{equation}
Static displacement, $\mu/2M\omega_0^2$, 
can be neglected compared to the oscillating part. 
Substituting  $X(t)$ into  Eq. (\ref{urabi}),
we find $u(t)$
\begin{equation}
\label{sol for u}
u(t) =  {-}\frac{\mu^2\left(1-\cos 2st\right)}{8M\Omega_R\bigl(\omega_0^2 -s^2\bigr)}.
\end{equation}
Substituting Eqs. (\ref{sol rabi x}), (\ref{sol for u})
into the right-hand side of Eq. (\ref{wrabi}), and equating 
the terms $\propto \cos st $ in both sides, we  find a closed 
equation for  $s$
\begin{equation}
\label{firstcubic}
\Omega_R^2 -s^2=\frac{\omega_p^2 \omega_0^4}{8\bigl(\omega_0^2-s^2\bigr)^2}.
\end{equation}
Thus, coupling to the oscillator causes the shift of the oscillation
frequency from $\Omega_R$, as stated in the Introduction.
Note that the term $\propto\cos 2st$ in $u(t)$ will also give rise to 
nonresonant contribution $\propto\cos 3st$ 
in $w(t)$, causing a weak anharmonicity of the oscillations.
Away from resonance, we can substitute $s=\Omega_R$ into the right-hand side
of Eq. (\ref{firstcubic}).  
Then Eq. (\ref{firstcubic}) yields a correction 
to the Rabi frequency due to coupling
to the oscillator
\begin{equation}
\label{s general}
s =\Omega_R-\frac{\omega_p^2 \omega_0}{64\bigl(\omega_0-\Omega_R\bigr)^2}.
\end{equation}
\begin{figure}[t]
	\begin{center}
	\includegraphics[width=50mm, angle=0,clip]{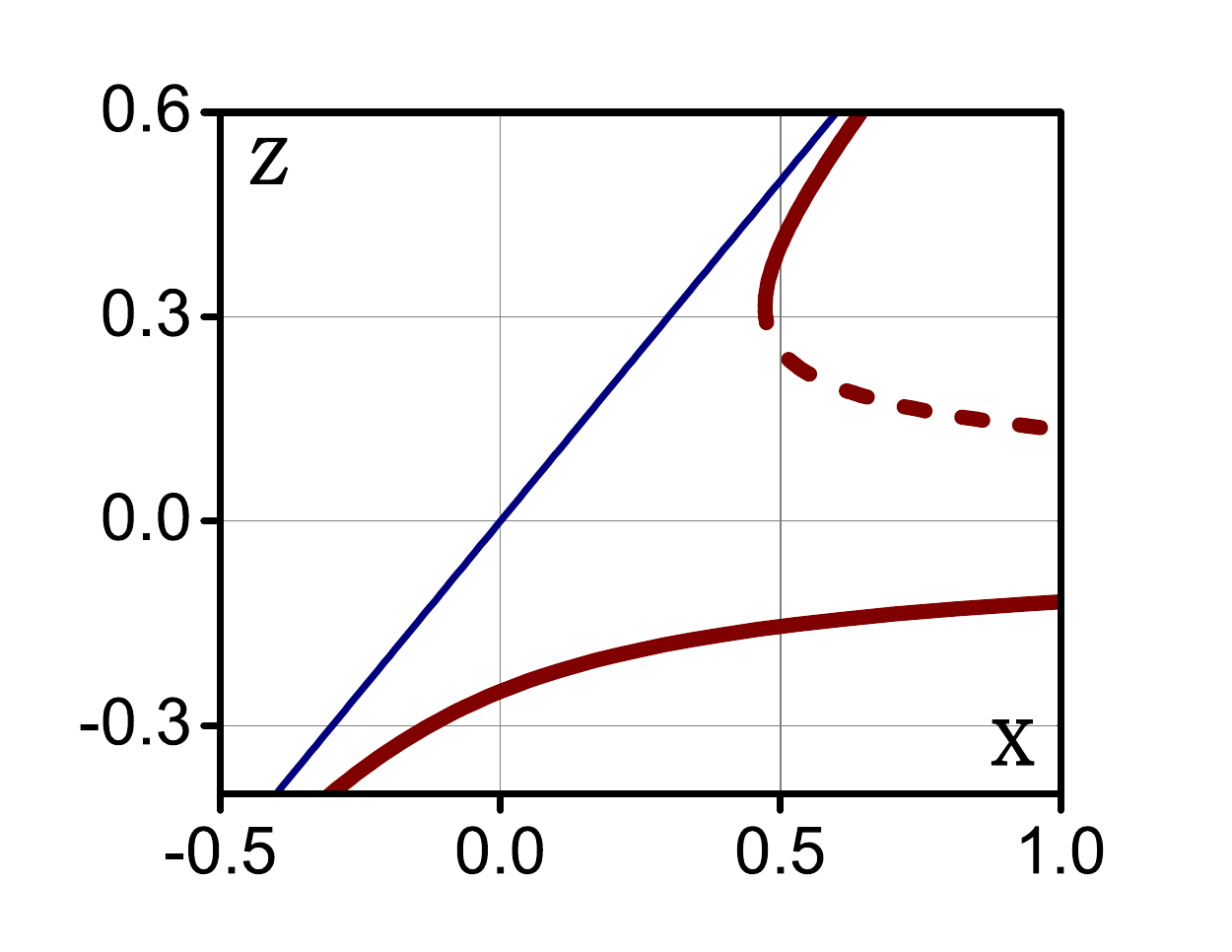}
	\end{center}	
	\caption{Red line: Dimensionless frequency, $z$, defined by Eq. (\ref{z}), of 		
	Eq. (\ref{dimensionless}) versus the dimensionless deviation, $x$, from the 		
	$\Omega_R = \omega_0$. Unstable solution is shown with dashed line. Blue 		line, 
	corresponds to the absence of coupling  to the oscillator.}
	\label{weakcoupling}
\end{figure}
This expression is valid only if the correction on the right-hand side is much smaller
than $\bigl(\omega_0-\Omega_R\bigr)$. Equating the correction to
$\bigl(\omega_0-\Omega_R\bigr)$, we find that the width 
of the resonance, ${\bigl(\omega_0-\Omega_R\bigr) \sim} \delta_0$, is given by Eq. (\ref{delta0}).

Recall now  our basic assumption that the oscillator is classical. We are now in position
to verify this assumption. In the resonant domain the amplitude, $X(t)$, can be estimated
from Eq. (\ref{sol rabi x}) as $X\sim \mu/M\omega_0\delta_0$. Then for the ratio of 
the energy of oscillations to the vibrational quantum, $\omega_0$, we get
the following estimate  
\begin{equation}
\label{estimate}
\frac{M X^2 \omega_0^2}{\omega_0} \sim  \left ( \frac{\omega_0}{\omega_p} \right )^{1/3}=\lambda^{-2/3} \gg 1.
\end{equation}
Thus, for weak coupling, the classical treatment of the oscillator is justified.

\subsection{Vicinity of the resonance}
To incorporate finite detuning, $\Delta$, into Eq. (\ref{firstcubic}) it is convenient
to rewrite Eq. (\ref{wrabi}) keeping all $\Delta$-dependent terms in the
right-hand side
\begin{align}
\label{withDelta}
\ddot{w} \hspace{1mm}+&\hspace{1mm}\Omega_R^2w+\mu^2X(t)\int_0^t dt^{\prime}X(t^{\prime})\dot{w}(t^{\prime})- \Delta \mu X(t) + \Delta^2\nonumber \\
& 
=\Delta\bigl(\mu X(t)-\Delta\bigr)w(t)+\Delta \mu\int_0^tdt^{\prime} X(t^{\prime})\dot{w}(t^{\prime}).\quad
\end{align}
The term $\propto \Delta^2$ in the right-hand side leads to a standard modification of the
Rabi frequency to $\bigl(\Omega_R^2+\Delta^2\bigr)^{1/2}$.
The last term is proportional to $\sin^2 st$, and does not contain the first harmonics.
The term $\propto \cos st$ comes from the combination $\Delta \mu X(t)$ in the left-hand side.
Emergence of this term, which is odd in detuning, is the result of the coupling of the
vibronic mode only to the level $E_1$. This term results in the 
following modification of Eq. (\ref{firstcubic})
\begin{equation}
 \label{Equationfors}
\Omega_R^2+\Delta^2 -s^2 = \frac{\omega_p^2 \omega_0^4}{{ 8}\bigl(\omega_0^2-s^2\bigr)^2} -
 \frac{\omega_p \omega_0^2\Delta}{\omega_0^2-s^2}.
\end{equation}
Near the resonance $\bigl(\Omega_R-\omega_0\bigr)\ll \omega_0$ this equation
can be simplified. Upon introducing dimensionless variables,
\begin{equation}
\label{z}
z=\frac{s- \omega_0 }{ \omega_p^{2/3}\omega_0^{1/3}},
\end{equation}
\begin{equation}
\label{x}
x= x'+ 8\Delta'^2,~~~x'=\frac{\Omega_R - \omega_0}{ \omega_p^{2/3} \omega_0^{1/3}},
\end{equation}
Eq. (\ref{Equationfors}) assumes the form
\begin{equation}
\label{dimensionless}
(z-x) z^2 + \Delta' z= -\frac{1}{64} ,
\end{equation}
where  dimensionless detuning, $\Delta'$, is defined as
\begin{equation}
\label{dprime}
\Delta'=\frac{\Delta} {4\omega_p^{1/3} \omega_0^{2/3}}.
\end{equation}
Note that characteristic detuning $\Delta \sim \omega_p^{1/3}\omega_0^{2/3}$ is much 
bigger than the width of the resonance, $\delta_0$, 
but much smaller than $\Omega_R$.
\begin{figure}
\begin{center}
\includegraphics[width=95mm, angle=0,clip]{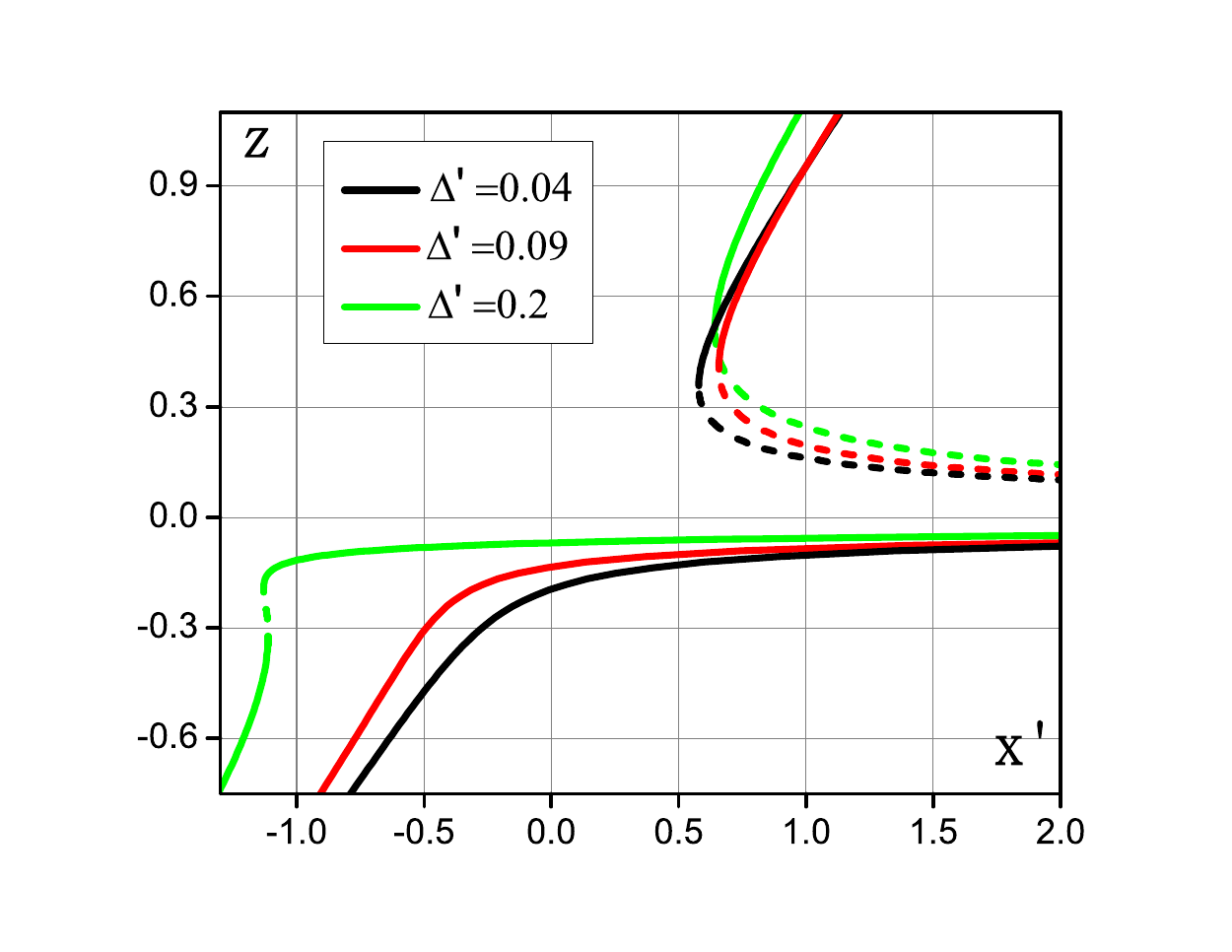}
\end{center}
\caption{Dimensionless frequency, $z$, of oscillations of driven two-level system 
is plotted from Eq. (\ref{dimensionless}) versus the dimensionless deviation, 
$x'$, from the resonance for three positive dimensionless detunings, 
$\Delta^{\prime}$, defined by Eq. (\ref{dprime}).  
As detuning  increases, the unstable branch shifts from positive $z$ to negative 
$z$, and both stable values of $z$ become positive (for positive $x'$) or negative 
(for negative $x'$).}
\label{detunepos}
\end{figure}
\begin{figure}
\begin{center}
\includegraphics[width=95mm, angle=0,clip]{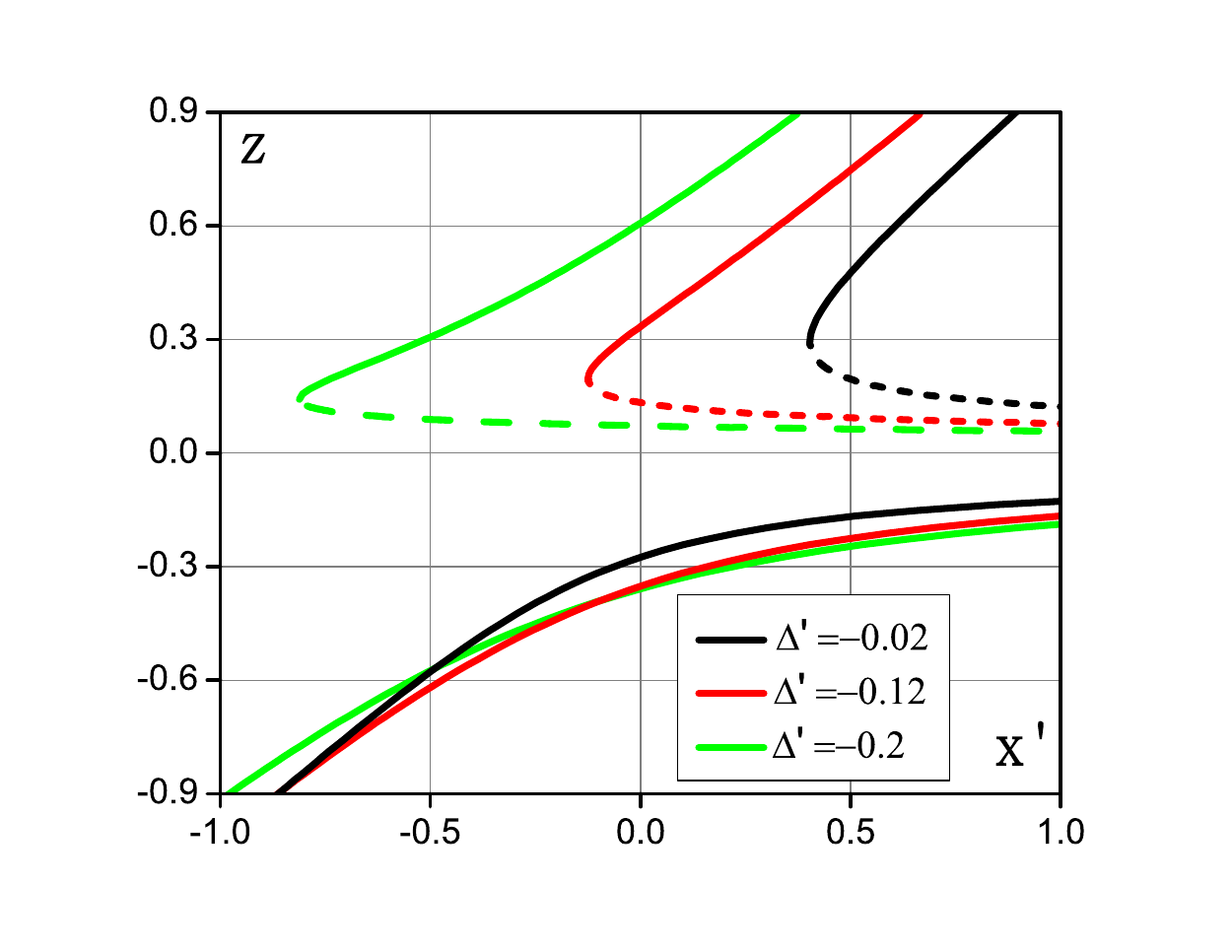}
\end{center}
\caption{ The same as in Fig.~\ref{detunepos} for three negative detunings. Note 
that negative $\Delta^{\prime}$ broadens the range of bistability.}
\label{detuneneg}
\end{figure}
Solution of Eq. (\ref{dimensionless}) for zero detuning is plotted 
in Fig.~\ref{weakcoupling}. Blue line $z=x$ corresponds to the Rabi 
oscillations without coupling. 
We see that bistability develops for $x> 3\cdot 2^{-8/3}$. 
At $x=3\cdot 2^{-8/3}$, the frequency $s$ experiences a jump by $2^{-8/3}\delta_0$. 
Two values of $z$ corresponding 
to stable solutions define, via Eq. (\ref{z}),  two frequencies of the 
modified Rabi oscillations. They also define two corresponding amplitudes 
of the oscillator
\begin{equation}
\label{amplitudeX}
X = \Bigl(\frac{1}{{2}\lambda^{1/3}z}\Bigr)\left (2M\omega_0\right)^{-1/2}\hspace{-.7cm}.
\end{equation}
The last factor in Eq. (\ref{amplitudeX}) is the amplitude
of a zero-point motion of the oscillator.
As the dimensionless deviation, $x$, from resonance increases, 
the upper branch approaches $z=x$. For this branch the frequency 
of the Rabi oscillations is close to $\Omega_R$ and the amplitude 
of oscillator is small. For the lower branch $z$ is small, i.e., 
the frequency of the oscillations approaches $\omega_0$ with increasing 
$x$. For this branch the oscillator is highly excited. 

 Figs. 3 and  4 illustrate the effect of detuning on the frequency of oscillations, $s$.
 Note that there is a qualitative  difference between Fig. 2  for zero detuning, and
 Figs. 3 and 4 for positive and negative detunings, respectively. For zero detuning,
the domain of bistability exists only when $\Omega_R > \omega_0$, whereas for finite
detuning, bistable regions emerge both to the left and the right from the resonance.
As one changes the dimensionless deviation, $x'$,  from the resonance, from negative
to positive, for $\Delta'=0$,  bistability corresponds to $x'> 3\cdot 2^{-8/3}$.
For finite positive detuning, $\Delta' >0$, the first domain of bistability
occurs at $x'<0$, then disappears, and reemerges at positive $x'$ greater than $3\cdot 2^{-8/3}$. Conversely,
finite negative detuning simply broadens the domain of bistability as compared to $\Delta =0$.
Bistable region starts for $x'<0$. Peculiar dependence of $s$ on the deviation
 from resonance is also reflected in the amplitude of the oscillator.
This effect is discussed in the Sect. V.
\subsection{Effect of intrinsic anharmonicity of the oscillator}
Suppose that in addition to the harmonic part of the oscillator energy, $M\omega_0^2 X^2/2$, 
a weak intrinsic anharmonicity, $\kappa X^4/4$, is present. Then Eq. (\ref{sol rabi x}) will
assume the form 
\begin{equation}
\label{X_a}
\frac{3}{4}\kappa X_a^3+\bigl(\omega_0^2-s^2\bigr)X_a=\frac{\mu}{2M}.
\end{equation} 
The second relation, $s^2=\Omega_R^2-\mu^2X_a^2/{8}$, between the amplitude, 
$X_a$, and the frequency, $s$, which follows from Eq. (\ref{wrabi}) 
remains unchanged. It is now more convenient to express $s$ from
this relation and substitute it into Eq. (\ref{X_a}). This yields a
cubic equation for $X_a$
\begin{equation}
\label{Acubic}
2\omega_0\bigl(\omega_0-\Omega_R\bigr)X_a+\frac{6\kappa+\mu^2}{8}X_a^3=\frac{\mu}{2M}.
\end{equation}
If we now set $\kappa =0$,  then 
Eq. (\ref{Acubic}) 
will have multiple real solutions for $X_a$ 
in the domain $\bigl(\Omega_R-\omega_0\bigr)<-3\cdot 2^{-8/3}\delta_0$,
i.e., the same as determined from Eq. (\ref{dimensionless}) with $\Delta'=0$. We see from Eq. (\ref{Acubic}) 
that, depending on the sign of $\kappa$, intrinsic anharmonicity can 
either shift the threshold of bistability to the left 
(for positive $\kappa$) or to the right
(for negative $\kappa$).  Anharmonicity  will also affect the  
magnitude of the jump of the frequency, $s$, of the  oscillations. This magnitude will get
modified from $2^{-8/3}\delta_0$ to 
$2^{-8/3}\delta_0\bigl(1+6\kappa/\mu^2\bigr)^{-2/3}$, i.e., the jump will
become smaller for $\kappa <0$.

\section{Decay of the oscillations}
Up to now we disregarded both mechanisms of dissipation: finite relaxation, $\Gamma$, and the friction in the oscillator,
 $\gamma$. Rabi oscillations will decay with the rate $\Upsilon$, which is determined by $\Gamma$, in the
 regime $\Gamma \gg \gamma$, or by friction in the
 regime $\Gamma \ll \gamma$. We will consider both cases separately.
We emphasize that, as the oscillations decay, so does the coupling between the oscillator 
and two-level system. Thus the decay will be accompanied by the change of frequency 
back to $\Omega_R$. We cannot capture this  evolution of frequency  with time analytically. 
To find the decay rate, $ \Upsilon$, only, we will adopt the approach based on the 
energy conservation.

\subsection{Friction-dominated regime, $\gamma \gg \Gamma$}
Upon neglecting $\Gamma$ in Eq. (\ref{wrabi}) and setting $\Delta =0$ we have
\begin{equation}
\label{w conservation}
\ddot{w} + \Omega_R^2 w = - \mu^2 X(t) \int_0^t dt' X(t') \dot{w}(t').
\end{equation}
Multiplying both sides by $\dot{w}$ and integrating from $0$ to $t$,
 we arrive to the following conservation law
\begin{equation}
\label{friction conservation}
\frac{\dot{w}^2}{2} + \frac{\Omega_R^2}{2}(w^2 -1) = -\frac{\mu^2}{2}\left ( \int_0^t dt' X(t') \dot{w}(t') \right )^2\hspace{-2.3mm}.
\end{equation}
The right-hand side describes the energy exchange between the 
two-level system and the oscillator.

As a next step we multiply the equation of motion of the oscillator
Eq. (\ref{EOMx}), by $\dot{X}$  and integrate from $0$ to $t$. 
Then we arrive at the second conservation law
\begin{equation}
\label{x conservation}
\frac{\dot{X}^2}{2} + \frac{\omega_0^2 X^2}{2} +\gamma \int_0 ^t dt' \dot{X}^2(t') =- \frac{\mu}{2M} \int_0^t dt' w(t')\dot{X}(t').
\end{equation}
At long time, $t \gg \Upsilon^{-1}$ we have, $\dot{w}$, $w \rightarrow 0$ and also $\dot{X}$, $X \rightarrow 0$. Then the
left-hand side of Eq. (\ref{friction conservation}) for $w$ turns to $ -\Omega_R^2/2$. Combining Eqs.  (\ref{friction conservation})
and  (\ref{x conservation}),  we arrive to the relation
\begin{equation}
\Omega_R ={ 2} M \gamma  \int_0^\infty dt' \dot{X}^2(t').
\end{equation}
This relation is convenient to find the decay rate, $\Upsilon$, because it contains only $\dot{X}^2$, which is
insensitive to the change of frequency in the course of decay. Substituting $\dot{X}^2 \propto \exp[-2\Upsilon t]$,
we find the following relation between $\Upsilon$ and the amplitude of the oscillations at time $\lesssim 1/ \Upsilon$
\begin{equation}
\label{upsilon}
\Upsilon = \gamma (M X^2 \Omega_R).
\end{equation}
Note that the second factor in the right-hand side can be rewritten as $ (M X^2 \Omega_R^2)/ \Omega_R$.
The oscillator is classical when this ratio is big. Thus we conclude 
that $\Upsilon \gg \gamma$, i.e., the Rabi
oscillations in the region of resonance decay faster than undriven oscillator.
 As a next step, we distinguish two
cases of weak and strong friction. 
In the first case, to find $X$ that should be substituted into Eq. (\ref{upsilon}) 
one can use Eq. (\ref{sol rabi x})
 obtained without friction. Then one gets
\begin{equation}
\label{upsilon general}
\Upsilon = \frac{\gamma \omega_p \Omega_R}{(\omega_0 - s)^2},
\end{equation}
where $s$ is determined from the cubic equation Eq (\ref{s general}). In the region of resonance, 
the difference $\omega_0-s$ is $\sim \delta_0$, which yields
\begin{equation}
\label{Upsilon small gamma limit}
\Upsilon \sim \gamma \left ( \frac{\omega_0}{\omega_p}\right )^{1/3}\hspace{-5
mm}.
\end{equation}
Weak friction requires that $|\omega_0 -s | \sim \delta_0 \gg \Upsilon $, i.e.,  $\gamma \ll \omega_p$.

 In the region of strong friction the difference $\omega_0 - s$ should be replaced by $\Upsilon$. 
Then Eq. (\ref{upsilon general}) contains $\Upsilon$ in both sides. 
Upon solving this equation, we get 
\begin{equation}
\label{Upsilon large gamma limit}
\Upsilon \sim \left (\frac{\gamma}{ \omega_p}\right)^{1/3} \delta_0,
\end{equation}
for $\gamma \gg \omega_p$. Equations (\ref{Upsilon small gamma limit}), 
(\ref{Upsilon large gamma limit}) match when $\gamma \sim \omega_p$. 
The validity of this expression is limited from above by the condition that the 
oscillator is classical. As we replace $\omega_0 -s$ by $ \Upsilon$, the 
estimate for $X$ is $X\sim \mu /(M \omega_0 \Upsilon)$. 
Then the kinetic energy can be estimated as 
$M \omega_0^2\bigl[\mu /(M \omega_0 \Upsilon)\bigr]^2 \sim \omega_p^{1/3} \omega_0^{4/3} / \gamma^{2/3}$.
 The condition that it is bigger than $\omega_0$ limits $\gamma$ in 
Eq. (\ref{Upsilon large gamma limit}) to
$\gamma \ll \delta_0(\omega_0 / \omega_p)^{1/6}$, and correspondingly limits $\Upsilon$
to $\Upsilon \ll \delta_0(\omega_0 /\omega_p)^{1/6}$. 

From Eqs.   (\ref{Upsilon small gamma limit}), (\ref{Upsilon large gamma limit}) we
see that, upon increasing friction, the decay rate, $\Upsilon$, first grows linearly
with $\gamma$, and then sublinearly, as $\gamma^{1/3}$. At the boundary of
applicability of the classical description we have $\Upsilon=\gamma$.  
For even bigger $\gamma$ classical treatment of the oscillator is not justified, 
but we expect that
the oscillator  will eventually decouple from the two-level system, 
and Rabi oscillations will proceed as they do in the absence of the oscillator.

 \subsection{Initial stage of the oscillations}
After the ac driving field is switched on, the 
population inversion starts to
oscillate with frequency $\Omega_R$. 
After some  number, $m$,  of
the  oscillation cycles the frequency crosses over to $s$. 
The question of interest is how $m$ depends on the coupling 
strength, $\lambda$.  We will estimate $m$ using the fact 
that at initial stage 
the system Eqs. (\ref{EOMx}), (\ref{w conservation}) 
can be solved perturbatively in small parameter $\omega_p/\omega_0$.
To find the perturbative solution, we substitute
the "bare" Rabi oscillations $w^{(0)}=-\cos(\Omega_Rt)$ into Eq. (\ref{EOMx}) and
find $X(t)$ with initial conditions $X(0)=0$, $\dot{X}(0)=0$. 
The obtained $X(t)$ together with $w^{(0)}(t)$ is then 
substituted into the right-hand side
of  Eq. (\ref {w conservation}). Solving this second-order 
differential equation with a given right-hand side, 
we find that the amplitude of oscillations    becomes $1-w^{(1)}(t)$, i.e.,
it acquires a time-dependent correction with $w^{(1)}(t)$ given by the
following expression
\begin{eqnarray}
\label{correction}
w^{(1)}(t)&=& \frac{2 \omega_p^2 \omega_0^2}{(\gamma^2 + 4\delta^2)^2} \bigg [ \frac{\gamma t}{2} - \frac{4 \gamma \delta}{ (\gamma^2 + \delta^2)} e^{-\gamma t/2} \sin \delta t\nonumber \\
&-& \frac{\gamma^2 -4\delta^2}{\gamma^2 + 4\delta^2}\left ( 1 - e^{-\gamma t/2} \cos \delta t \right )  \bigg ]^2\hspace{-1.5mm},
\end{eqnarray}
where $\delta= \omega_0-\Omega_R$. We can now estimate $m$ as 
$\Omega_Rt_c\approx \omega_0t_c$, where $t_c$ is the
time after which $w^{(1)}(t)$ becomes $\sim 1$. Consider first the 
limit  $\delta \rightarrow 0$.
Then $w^{(1)}(t)$ grows with time as $\bigl (\omega_p\omega_0t/\gamma\bigr)^2$. 
This yields $m \sim \gamma/\omega_0\lambda^2$. 
Condition that $m\gg 1$ should be consistent with the
condition that the oscillator is classical. 
The latter condition reads: $\omega_p\omega_0\gg \gamma^2$.
In the domain when both conditions are met 
we have $1\ll m \ll \frac{1}{\lambda}$. 
 
Consider now the limit $\gamma\rightarrow 0$. 
For $\delta t \ll 1$, 
Eq. (\ref{correction}) yields 
$w^{(1)}(t) \sim \left(\omega_p\omega_0t^2\right)^2$.
This leads to the estimate $t_c \sim \bigl(\omega_p\omega_0\bigr)^{-1/2}$ and
$m \sim  \lambda^{-1}$. Small-$t$ expansion of Eq. (\ref{correction}) is valid if
the product $\delta t_c$ is small. With $t_c$ found above, this product can rewritten
in the form 
\begin{equation}
\label{t_c}
\delta t_c \sim \biggl(\frac{\delta^2}{\omega_p\omega_0}\biggr)^{1/2}\hspace{-.5cm}.
\end{equation}
On the other hand, the oscillator can  be treated as classical when the ratio
in the right-hand side of Eq. (\ref{t_c}) is small. Thus,  taking the limit $\delta t \ll 1$
in  Eq. (\ref{correction}) is justified, and the frequency of the Rabi oscillations crosses
over from $\Omega_R$ to $s$ after $\sim \lambda^{-1}$ cycles. Correspondingly,
after $\sim\lambda^{-1}$ cycles, 
the oscillator will "forget" about initial phase, imposed by the initial conditions, and
will execute a forced harmonic motion with frequency, $s$.

\subsection{Relaxation-dominated regime, $\Gamma \gg \gamma$}
At finite relaxation rate of the two-level system Eq. (\ref{x conservation}) assumes the form
\begin{widetext}
\begin{equation}
\label{Generalized}
\frac{\dot{w}^2}{2} + \frac{1}{2}\left (\Omega_R^2+ \frac{\Gamma^2}{2} \right ) \left ( w^2 -1 \right) + \frac{\Gamma^2}{2}
\left(w+1\right ) +\frac{3}{2} \Gamma \hspace{-.5mm}\int_0^t \hspace{-1.5mm}dt' \dot{w}^2= 
-\mu^2 \hspace{-.5mm}\int_0^t\hspace{-1.5mm} dt'X(t')\dot{w}
(t')\hspace{-.5mm} 
\int_0^{t'}\hspace{-1.5mm}dt'' e^{\Gamma(t''-t')/2} X(t'') \left [ \dot{w}(t'') + \Gamma (1+ w) \right ]\!.
\end{equation}
\end{widetext}
Without coupling to the oscillator the right-hand side is zero, and Eq. (\ref{Generalized}) describes the 
decay of the Rabi oscillations due to relaxation. Indeed, upon substituting $\dot{w}=
\Omega_R\sin\bigl(\Omega_Rt'\bigr)\exp(-\Upsilon t')$ and taking the limit $t\rightarrow \infty$,
the last term in the left-hand side takes the value $3\Gamma\Omega_R^2/8\Upsilon$, which  leads to
$\Upsilon=3\Gamma/4$. Naturally, this value of $\Upsilon$ follows 
directly from Eqs. (\ref{wdensity}), (\ref{vdensity}). Finite coupling to the oscillator would increase the decay rate only if 
at $t\rightarrow \infty$ the integral in the right-hand side  exceeds $\Omega_R^2$. 
Contribution of the second term in the square brackets to the integral can be estimated
upon noticing that the product $X(t'')w(t'')$ is a slow function. 
Assuming that $X$ and $w$ both
decay as $\exp(-\Upsilon t'')$ and that $\Upsilon \gg \Gamma$,
the integral over $t'$ reduces to 
$\int_0^\infty dt't'\sin(2st')\exp(-2\Upsilon t')=\Upsilon/s^3\approx \Upsilon/\omega_0^3$.
Then one gets the estimate
$\omega_p^2\Upsilon\Gamma/\bigl(\omega_0-s\bigr)^2$ for this contribution. Since 
$\omega_0-s$ cannot be smaller than $\Upsilon$, this contribution cannot exceed 
$\omega_p^2$ which is much smaller than $\Omega_R^2$. The contribution from 
the first term in the square brackets also cannot exceed $\Omega_R^2$. 
This becomes apparent upon performing integration by parts after  
which the contribution from the first term assumes the form 
\begin{equation}
\label{by parts}
-{\Gamma}\mu^2\int_0^{\infty} dt'e^{-\Gamma t'} \left( \int_0^{t'} dt''  e^{\Gamma t''/2} X(t'') \dot{w}(t'')\right )^2 \hspace{-2.1mm}.
\end{equation}
If $X$ and $\dot{w}$ decay much faster than $\exp(-\Gamma t''/4)$, the inner integral saturates
at times $t' \ll \Gamma$. Then the contribution Eq. (\ref{by parts}) can be estimated as 
$\Omega_R^2\omega_p^2/\bigl(\omega_0-s\bigr)^2$, which is again much smaller
than $\Omega_R^2$. We thus conclude that, while coupling to the oscillator modifies the
frequency, the decay of the Rabi oscillations in the relaxation-dominated regime   
is always dominated by the relaxation rate.

\section{Number of vibrational quanta}
\begin{figure}
\begin{center}
\includegraphics[scale=.3]{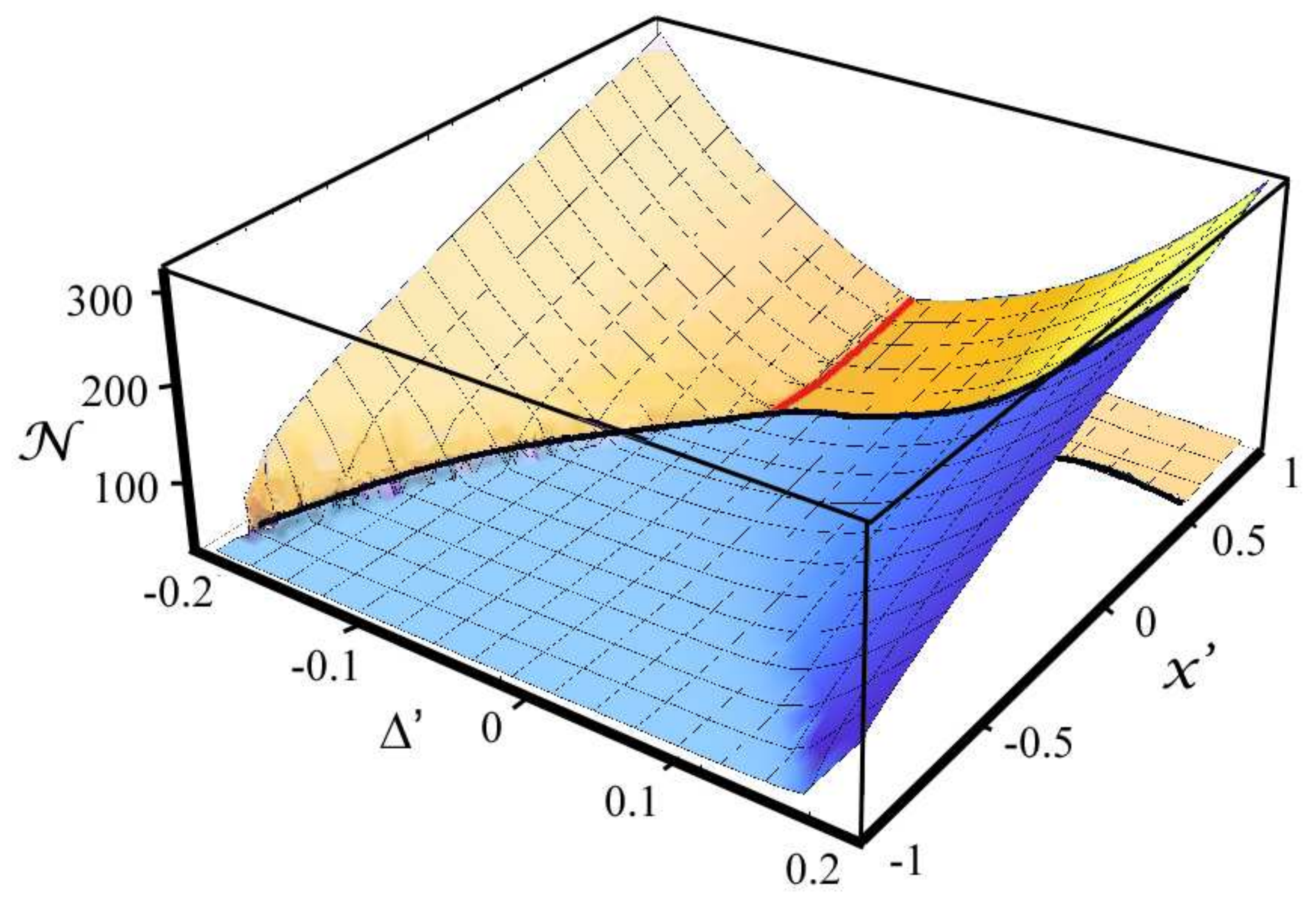}
\end{center}
\caption{ Excitation level of the oscillator, $\mathcal N = 16N  \lambda^{2/3}$, 
where $N$ is the number of vibrational quanta, is plotted from Eq. 
(\ref{dimensionless}) vs. dimensionless detuning and dimensionless deviation
 from the  resonance, $x'= \big( \Omega_R-\omega_0)/(\omega_0 \omega_p^2 \big )^{1/3}$. 
The thick black line of bifurcations separates the "inner" domain of parameters 
(blue domain where bistability is absent) and outer domain (yellow domain where 
bistability is present). In the outer domain the ratio of $\mathcal N$- values 
corresponding to the two stable solutions grows rapidly away from the boundary.}
\label{3dplot}
\end{figure}

We studied the behavior of the  of frequency of the Rabi oscillations near the  the resonance $\Omega_R=\omega_0$.
The number of vibrational quanta, $N$,  is also sensitive to the deviation, $\Omega_R-\omega_0$, from the resonance 
 and to the detuning, $\Delta$.  This number can be expressed  as follows
\begin{equation}
\label{N1}
N=  \left (\frac{1}{\lambda^{2}}  \right )^{1/3} \frac{1}{16 z^2} = \left ( \frac{ \omega_0}{\omega_p} \right)^{1/3} \frac{1}{16z^2},
\end{equation}
where $z$ is the solution of Eq. (\ref{dimensionless}).
The dependence of $N$ on dimensionless deviation, $x'$, and dimensionless detuning, $\Delta'$, is plotted in Fig.~\ref{3dplot}.
The values of $N$ shown correspond only to stable regimes of oscillations. The line of bifurcation points separates the ($x', \Delta^{\prime}$) domains with and without bistability in Fig.~\ref{3dplot}. In the domain of bistability the higher and lower values of $N$ coincide along the red line. 
Away from the red line the high-$N$ and the low-$N$ values differ very strongly. High-$N$ values correspond to the regime of oscillations with frequency close to $\omega_0$, see Fig.~\ref{detunepos}, whereas low $N$-values correspond to the frequency of oscillations close to $\Omega_R$. 

At $x'<0.5$  and to the right from the bifurcation line there is no bistability. The value of $N$ is low in this domain around $\Delta'=0$. As $\Delta'$ increases, $N$ grows for both
signs of detuning, $\Delta'$. However, for $\Delta'>0$ (blue detuning) the growth of $N$ is monotonical. At the same time, for $\Delta'<0$ (red detuning)
the bistability sets in at certain critical $\Delta'$. Upon further increase of $\vert \Delta'\vert$ the low-$N$ value does not grow, while the high-$N$ value 
grows rapidly.

It is instructive to compare the results shown in Fig.~\ref{3dplot} to the results of Refs. [\onlinecite{Shnirman, Shnirman1}]. The curves $N(\Delta',\Omega_R)$ were obtained in Refs. [\onlinecite{Shnirman, Shnirman1}] by numerical solution of the system of master equations for the density matrix describing
both the two-level system and the oscillator.  Firstly, there is a qualitative agreement in the shape of the boundary of bistability.  In Refs. [\onlinecite{Shnirman, Shnirman1}] only the low-$N$ values are plotted. The prime observation made in Refs. [\onlinecite{Shnirman, Shnirman1}]
was that there is a strong difference between these low-$N$ values for blue and red detunings, 
namely, for blue detuning $N$ is much higher. Our analytical results in Fig. \ref{3dplot} 
agree qualitatively with this  observation.

\section{Concluding remarks}
Frequency, $\Omega_R$,  of the  Rabi oscillations 
is proportional to the square root of the excitation
power. This linearity has been
demonstrated in many experiments.
Even when Rabi oscillations are
damped, the dependence $s(\Omega_R)$
can be extracted from 
the position of maximum in the Fourier
transform\cite{Boehme} of the signal, $w(t)$.
We predict that, for a two-level system coupled
to a vibrational mode, the position of maximum
of the Fourier transform will deviate from the
linear behavior near the resonance 
$\Omega_R=\omega_0$.
Both to the left and to the right from the resonance
 the position of maxim corresponds 
to $s<\Omega_R$.  The relative width of the
resonant region depends on the coupling, $\lambda$, to
the vibrational mode as $\lambda^{4/3}$.
We also predict that, in the vicinity of the resonance,
the dependence $s(\Omega_R)$ exhibits a hysteretic 
behavior with two stable values of $s$ corresponding to
two stable regimes of the Rabi oscillations. 

The underlying physics of the Rabi-vibronic resonance is the following. 
Without coupling, population inversion, $w$, and displacement, $X$,
satisfy the harmonic oscillator equations with frequencies $\Omega_R$ and
$\omega_0$, respectively. With coupling, two-level system acts as a driving
force $\propto w$ on the oscillator, while the back-action of the oscillator 
on $w$ is peculiar. The structure of back-action force is $wX^2$, as can be seen from
Eqs. (\ref{wrabi}), (\ref{urabi}). This structure implies that back-action is of
a parametrical type, i.e., $X^2$ adds to $\Omega_R^2$.
Thus, at $\Omega_R\approx \omega_0$,  it appears that $\Omega_R$ is modulated
with frequency $\approx 2\Omega_R$. This, however, does
not lead to a parametric  instability.  Instead, the oscillator motion gets 
 synchronized with the Rabi oscillations.
In this regard, there is certain
analogy to the synchronization of the Rabi
oscillations to a sequence of pulses \cite{korotkov} 
applied to the detector with 
repetition period chosen to be  $2\pi/\Omega_R$.

As it was pointed out in Introduction, the situation when 
a two-level system undergoing the Rabi oscillations is 
coupled to the oscillator is actively studied in connection to
the circuit QED\cite{QEDGirvin}. The most common situation
in circuit QED is when the oscillator frequency, $\omega_{0}$,
is tuned to the transition frequency, $\omega_{12}$, of the
two-level system.   Among physical effects predicted for
this domain is that two or multiple qubits can get strongly
coupled  to each other via coupling to a common oscillator\cite{Koch,Blanter}.  
Rabi-vibronic resonance corresponds to the domain $\omega_0 \ll \omega_{12}$.
Still the effects similar to those discussed in Refs. [\onlinecite{Koch,Blanter}]
 will take place under the conditions of the Rabi-vibronic resonance.
In particular, we anticipate that Rabi oscillations in two driven 
two-level systems with $\Omega_R=\omega_0$,
 coupled to the same oscillator will get synchronized.

As a final remark, classical treatment of the vibrational mode adopted in the present
paper  does not allow one to capture the quantum 
jumps\cite{OnlyOscillatorDriven} between
the stable regimes of collective motion of the two-level system 
coupled to the oscillator.  We also did not consider the effect  of 
 thermal noise which leads to the activated switching\cite{dykman} 
between the steady regimes even within a classical description 
of the oscillator.
 
\begin{acknowledgements}
We are grateful to C. B\"{o}hme for a discussion which initiated this study. We thank J. Koch and F. von Oppen
for bringing Refs. [\onlinecite{Shnirman,Shnirman1}] to our attention. We also acknowledge discussions with E. G. Mishchenko, O. A. Starykh,  and B. Spivak, and the support of the Grant  DMR-0808842 (R.G).

\end{acknowledgements}

\end{document}